\documentclass[12pt]{article}
\usepackage{amsmath,amssymb}
\usepackage{bm}
\usepackage{color}
\usepackage{cite}
\usepackage{mathtools}

\usepackage{multirow}

\begin{document}

\title{
\begin{flushright}
\ \\*[-80pt]
\begin{minipage}{0.25\linewidth}
\normalsize
EPHOU-22-001\\
HUPD-2201\\
KEK-TH-2387\\
KYUSHU-HET-234 \\*[50pt]
\end{minipage}
\end{flushright}
{\Large \bf
4D modular flavor symmetric models \\ inspired  by higher dimensional theory
\\*[20pt]}}

\author{
Shota Kikuchi  $^{1}$,
~Tatsuo Kobayashi  $^{1}$, 
 Hajime Otsuka  $^{2,3}$, \\ Morimitsu Tanimoto $^{4}$,  ~Hikaru Uchida$^{1}$ and Kei Yamamoto  $^{5}$
\\*[20pt]
\centerline{
\begin{minipage}{\linewidth}
\begin{center}
$^1${\it \normalsize
Department of Physics, Hokkaido University, Sapporo 060-0810, Japan} \\*[5pt]
$^2${\it \normalsize 
KEK Theory Center, Institute of Particle and Nuclear Studies,}\\
{\it \normalsize 1-1 Oho, Tsukuba, Ibaraki 305-0801, Japan}\\
$^3${\it \normalsize 
Department of Physics, Kyushu University, 744 Motooka, Nishi-ku, Fukuoka,}\\
{\it \normalsize 819–0395, Japan}
\\*[5pt]
$^4${\it \normalsize
Department of Physics, Niigata University, Niigata 950-2181, Japan} \\*[5pt]
$^5${\it \normalsize
Core of Research for the Energetic Universe, Hiroshima University, Higashi-Hiroshima 739-8526, Japan} \\*[5pt]
\end{center}
\end{minipage}}
\\*[50pt]}

\date{
\centerline{\small \bf Abstract}
\begin{minipage}{0.9\linewidth}
\medskip
\medskip
\small
We study a scenario to derive four-dimensional modular flavor symmetric models 
from higher dimensional theory by assuming the compactification consistent with 
the modular symmetry.
In our scenario, wavefunctions in extra dimensional compact space 
are modular forms.
That leads to constraints on combinations between modular weights 
and $\Gamma_N$ ($\Gamma_N')$ representations of matter fields.
We also present illustrating examples.
\end{minipage}
}

\begin{titlepage}
\maketitle
\thispagestyle{empty}
\end{titlepage}

\newpage


\section{Introduction}
\label{Intro}

The supersymmetric (SUSY) modular invariant theories give us an attractive framework to address the flavor problem of quarks and leptons.
Indeed, finite modular flavor symmetric models have been presented for years 
\cite{Feruglio:2017spp,Kobayashi:2018vbk,
	Penedo:2018nmg,Novichkov:2018nkm,	Ding:2019xna,Liu:2019khw,Chen:2020udk,
	Novichkov:2020eep,Liu:2020akv,Wang:2020lxk,Yao:2020zml,Kobayashi:2018bff,Ding:2020msi}. 
The homogeneous modular group $\Gamma=SL(2,\mathbb{Z})$ 
and inhomogeneous modular group $\bar \Gamma=SL(2,\mathbb{Z})/\mathbb{Z}_2$ 
include $S_3, A_4, S_4, A_5$ as finite subgroups  \cite{deAdelhartToorop:2011re}.
Indeed, the quotients $\Gamma_N=\bar \Gamma/\Gamma(N)$  are isomorphic to 
$\Gamma_3\simeq A_4$, $\Gamma_4 \simeq S_4$, and $\Gamma_5 \simeq A_5$ while 
$\Gamma/\Gamma(2) \simeq S_3$, where 
$\Gamma(N)$ are principle congruence subgroups.
These non-Abelian flavor symmetries such as $S_3, A_4, S_4, A_5$ 
were often used to derive  quark and lepton mass matrices
successfully  in flavor model building before the studies of 
modular flavor models ~\cite{Altarelli:2010gt,Ishimori:2010au,Ishimori:2012zz,Hernandez:2012ra,King:2013eh,King:2014nza,Tanimoto:2015nfa,King:2017guk,Petcov:2017ggy,Feruglio:2019ktm}. 

In modular flavor models, Yukawa couplings  are 
modular forms depending on the modulus $\tau$, and are certain representations 
under $\Gamma_N$ and their covering groups $\Gamma_N'$.
We assign modular weights and $\Gamma_N$ ($\Gamma_N'$) representations 
to matter fields as well as Higgs fields, although 
Higgs fields are assigned to a $\Gamma_N$ ($\Gamma_N'$) trivial singlet in most of the modular flavor models.
Then, the structure of quark and lepton mass matrices is given by certain modular forms 
under the assumption that the Yukawa coupling terms (in the superpotential) 
as well as mass terms 
are invariant under the modular symmetry. 
By taking these modular flavor symmetric mass matrices, 
one can realize realistic quark and lepton masses and mixing angles
by fixing the modulus $\tau$.
The CP violation and  related phenomena have also been studied \cite{Criado:2018thu,Kobayashi:2018scp,Ding:2019zxk,Novichkov:2018ovf,Kobayashi:2019mna,Wang:2019ovr,
 	Okada:2020brs,Yao:2020qyy,Okada:2018yrn,Okada:2019uoy}.
 Besides mass matrices of quarks and leptons,
 related topics such as grand unified theory, leptogenesis, dark matter, etc., have been discussed in many works
\cite{deMedeirosVarzielas:2019cyj,
	Asaka:2019vev,Asaka:2020tmo,Behera:2020sfe,Mishra:2020gxg,deAnda:2018ecu,Kobayashi:2019rzp,Novichkov:2018yse,Kobayashi:2018wkl,Nomura:2019jxj, Okada:2019xqk,
	Kariyazono:2019ehj,Nomura:2019yft,Okada:2019lzv,Nomura:2019lnr,Criado:2019tzk,
	King:2019vhv,Gui-JunDing:2019wap,deMedeirosVarzielas:2020kji,Zhang:2019ngf,Nomura:2019xsb,Kobayashi:2019gtp,Lu:2019vgm,Wang:2019xbo,King:2020qaj,Abbas:2020qzc,Okada:2020oxh,Okada:2020dmb,Ding:2020yen,Nomura:2020opk,Nomura:2020cog,Okada:2020rjb,Okada:2020ukr,Nagao:2020azf,Nagao:2020snm,Abbas:2020vuy,Feruglio:2021dte,King:2021fhl,Chen:2021zty,Novichkov:2021evw,Du:2020ylx,Kobayashi:2021jqu,Ding:2021zbg,Kuranaga:2021ujd,Li:2021buv,Tanimoto:2021ehw,Okada:2021aoi,Kobayashi:2021ajl,Dasgupta:2021ggp,Nomura:2021ewm,Nagao:2021rio,Nomura:2021yjb,Nomura:2021aep,Okada:2021qdf,Ding:2021eva,Qu:2021jdy,Zhang:2021olk,Wang:2021mkw,Wang:2020dbp,Kikuchi:2021yog
}.
It is also remarked that the standard model effective field theory (SMEFT) has been studied in the modular symmetry 
\cite{Kobayashi:2021uam,Kobayashi:2021pav}. 
Theoretical investigations have also been proceeded \cite{Baur:2019kwi,Baur:2019iai,Nilles:2020nnc,Nilles:2020kgo,Nilles:2020tdp,Baur:2020jwc,Nilles:2020gvu,Kobayashi:2020hoc,Kobayashi:2020uaj,Ishiguro:2020tmo,Baur:2020yjl,Ding:2020zxw,Baur:2021bly,Liu:2021gwa}.
Various combinations of matter modular weights and $\Gamma_N$  ($\Gamma_N'$) 
representations 
have been studied in order to lead to phenomenologically interesting results.

On the other hand, the modular symmetry is the geometrical symmetry of compact spaces 
such  as $T^2$ and the orbifold $T^2/\mathbb{Z}_2$.
Thus, four-dimensional modular flavor symmetric models could be derived from a higher dimensional theory 
such as the superstring theory.
For example, flavor transformations under the modular symmetry 
were studied in heterotic orbifold models \cite{Ferrara:1989qb,Lerche:1989cs,Lauer:1990tm} and magnetized D-brane models \cite{Kobayashi:2018rad,Kobayashi:2018bff,Ohki:2020bpo,Kikuchi:2020frp,Kikuchi:2020nxn,
Kikuchi:2021ogn,Almumin:2021fbk}.
Furthermore, Calabi-Yau compactifications have many moduli, and they 
have larger geometrical symmetries, i.e., symplectic modular symmetries 
$Sp(g,\mathbb{Z})$ 
\cite{Strominger:1990pd,Candelas:1990pi,Ishiguro:2020nuf,Ishiguro:2021ccl}.

However, in most four-dimensional (4D) 
modular flavor models, their relations with a higher 
dimensional theory are not clear: How do 4D modular 
flavor symmetric models appear as a 4D low-energy effective field theory 
from a higher dimensional theory?
Our purpose in this paper is to propose a scenario to derive 4D modular flavor symmetric 
models from a higher dimensional theory.
We do not specify its compactification, but we assume generic compactification consistent with the modular symmetry.
We study the Kaluza-Klein (KK) decompositions in a modular-symmetric way. 
In this scenario, wavefunctions in extra dimensional compact space can be written by 
modular forms.
Such a scenario leads to constraints of 4D modular flavor symmetric models.
Modular weights and representations of matter fields are constrained.

This paper is organized as follows.
In section \ref{sec:modular}, we give a brief review of the modular symmetry 
and modular forms.
We also study the structure of $\Gamma(3)$ modular forms.
In section \ref{sec:high-D}, 
we study a scenario to derive 4D modular flavor symmetric models from a 
higher dimensional theory with modular symmetric compactification.
In section \ref{sec:model}, 
we study illustrating examples with $A_4$ modular flavor symmetry.
Section \ref{sec:conclusion} is our conclusion.
In Appendix \ref{appendix:A}, we show an example to project 
wavefunctions with $\Gamma_N$ reducible representations to an irreducible one.


\section{Modular symmetry and modular forms}
\label{sec:modular}

\subsection{Modular symmetry}

Here, we briefly review the modular symmetry and modular forms.
The $SL(2,\mathbb{Z})=\Gamma$ group is a group of the following $2 \times 2$ matrices:
\begin{align}\gamma = 
\begin{pmatrix}
a &b \\
c & d
\end{pmatrix},
\end{align}
where $a,b,c,d$ are integers and $ad-bc=1$.
The $SL(2,\mathbb{Z})$ group is generated by $S$ and $T$,
\begin{align}
S=
\begin{pmatrix}
0 & 1 \\ -1 & 0
\end{pmatrix}, 
\qquad 
T = 
\begin{pmatrix}
1 & 1\\ 0 & 1
\end{pmatrix}.
\end{align}
They satisfy the following algebraic relations:
\begin{align}
S^4=1, \quad (ST)^3=1.
\end{align}

The modulus $\tau$ transforms as
\begin{align}
\tau \to \gamma \tau = \frac{a \tau +b}{c \tau +d}\,,
\end{align}
under the modular symmetry.
The generators $S$ and $T$ satisfy the following algebraic relations 
on $\tau$:
\begin{align}
S^2=1, \quad (ST)^3=1,
\end{align}
i.e., $PSL(2,\mathbb{Z})=SL(2,Z)/\mathbb{Z}_2=\bar \Gamma$.

The modular forms are described by a holomorphic function $f_i(\tau)$, which transforms 
under the modular symmetry as 
\begin{align}
f_i(\gamma \tau)=(c \tau +d)^k\rho(\gamma)_{ij}f_j(\tau),
\end{align}
with $k$ and $\rho(\gamma)_{ij}$ being the modular weight and 
unitary matrices, respectively. 

Here, we introduce the principal congruence subgroups
\begin{align}
\begin{aligned}
\Gamma(N)= \left \{ 
\begin{pmatrix}
a & b  \\
c & d  
\end{pmatrix} \in SL(2,\mathbb{Z})~ ,
~~
\begin{pmatrix}
a & b  \\
c & d  
\end{pmatrix} =
\begin{pmatrix}
1 & 0  \\
0 & 1  
\end{pmatrix} ~~({\rm mod} N) \right \}
\end{aligned} .
\end{align}
The $\Gamma(N)$ modular forms satisfy 
\begin{align}
f_i(\gamma \tau)=(c \tau +d)^k f_i(\tau),
\end{align}
for $\gamma \in \Gamma(N)$.
Thus, the unitary matrices are representations of quotients 
$\Gamma_N = \bar \Gamma/\Gamma(N)$.
Interestingly, the quotients $\Gamma_N$ with $N=3,4,5$ 
are isomorphic to $A_4, S_4, A_5$, respectively.
In addition, $\Gamma_8$ and $\Gamma_{16}$ include $\Delta(96)$ and $\Delta(384)$ \cite{Kobayashi:2018bff}. 
These are finite modular subgroups including irreducible triplet representations.
Moreover, the quotient $\Gamma_2=\Gamma/\Gamma(2)$ is isomorphic to $S_3$.

Since $S^2=1$ in $\bar \Gamma$ on the modulus $\tau$, 
the modular weight $k$ must be even.
The dimensions $d_k(\Gamma(N))$ of modular forms of 
weights $k$ and levels $N$ are determined mathematically and 
shown in Table \ref{tab:form-dim}.
These modular forms are $d_k$ representations of $\Gamma_N$.
In general, they are reducible representations, and can be decomposed to 
irreducible representations as shown in the next subsection for $N=3$.

\begin{table}[h]
\centering
\begin{tabular}{|c|c|c|} \hline
$N$ & $d_k(\Gamma(N))$ &$\Gamma_N$  \\ \hline
2 & $k/2 +1$ & $S_3$ \\ \hline
3 & $k+1$ & $A_4$ \\ \hline
4 & $2k+1$ & $S_4$ \\ \hline
5 & $5k+1$ & $A_5$ \\ \hline
\end{tabular}
\caption{ Dimensions of modular forms of the level $N$ and weight $k$.}
\label{tab:form-dim}
\end{table}

The above modular forms can be extended to $\Gamma = SL(2,\mathbb{Z})$, 
which is the double covering group of $\bar \Gamma$.
For this group, the modular weights can be odd integers, 
and $\rho(\gamma)_{ij}$ are representations of the double covering groups of 
$\Gamma_N$, $\Gamma_N'$.
Furthermore, we can extend the double covering group of $\Gamma = SL(2,\mathbb{Z})$.
The modular weights can be half-integers, and 
$\rho(\gamma)_{ij}$ are representations of the double covering groups of $\Gamma_N'$.
For example, such modular forms with half-integers are obtained 
in magnetized D-brane models on $T^2$ and $T^2/\mathbb{Z}_2$ \cite{Kikuchi:2021ogn}.

\subsection{$\Gamma(3)$ modular forms}

Here, we show explicitly $\Gamma(3)$ modular forms and their $A_4$ representations.
The $A_4$ group has four irreducible representations, ${\bf 3}, {\bf 1}, {\bf 1}', {\bf 1}''$.
Their tensor products are obtained as 
\begin{align}
& {\bf 3}\times {\bf 3} = {\bf 3}_s + {\bf 3}_a + {\bf 1} + {\bf 1}' + {\bf 1}'', \notag \\
& {\bf 3} \times {\bf 1} ={\bf 3} \times {\bf 1}'={\bf 3} \times {\bf 1}'' ={\bf 3},
\end{align}
where ${\bf 3}_s$ and ${\bf 3}_a$ are symmetric and anti-symmetric, respectively, 
and 
\begin{align}
{\bf 1}_m \times {\bf 1}_n = {\bf 1}_\ell,
\end{align}
where $\ell =m+n$ (mod 3), ${\bf 1}_0={\bf 1}$, ${\bf 1}_1={\bf 1}'$, 
and ${\bf 1}_2={\bf 1}''$.

The $\Gamma(3)$ modular forms of weight $k=2$ have dimension $d_2=3$, 
and they are the $A_4$ triplet.
Their explicit forms are written by \cite{Feruglio:2017spp}
\begin{align}
&\begin{aligned}
{ Y^{\rm (2)}_{\bf 3}}(\tau)=
\begin{pmatrix}
Y_1(\tau)  \\
Y_2(\tau) \\
Y_3(\tau)
\end{pmatrix}\,,
\end{aligned}
\end{align}
\begin{eqnarray} 
\label{eq:Y-A4}
Y_1(\tau) &=& \frac{i}{2\pi}\left( \frac{\eta'(\tau/3)}{\eta(\tau/3)}  +\frac{\eta'((\tau +1)/3)}{\eta((\tau+1)/3)}  
+\frac{\eta'((\tau +2)/3)}{\eta((\tau+2)/3)} - \frac{27\eta'(3\tau)}{\eta(3\tau)}  \right), \nonumber \\
Y_2(\tau) &=& \frac{-i}{\pi}\left( \frac{\eta'(\tau/3)}{\eta(\tau/3)}  +\omega^2\frac{\eta'((\tau +1)/3)}{\eta((\tau+1)/3)}  
+\omega \frac{\eta'((\tau +2)/3)}{\eta((\tau+2)/3)}  \right) , \label{Yi} \\ 
Y_3(\tau) &=& \frac{-i}{\pi}\left( \frac{\eta'(\tau/3)}{\eta(\tau/3)}  +\omega\frac{\eta'((\tau +1)/3)}{\eta((\tau+1)/3)}  
+\omega^2 \frac{\eta'((\tau +2)/3)}{\eta((\tau+2)/3)}  \right)\, ,
\nonumber
\end{eqnarray}
where $\eta(\tau)$ is the Dedekind eta function, 
\begin{align}
\eta(\tau) = q^{1/24}\prod^\infty_{n=1} (1-q^n), \qquad q={\rm exp}~(2 \pi i \tau).
\end{align}

The modular forms of higher weights are obtained by the tensor products of 
${ Y^{\rm (2)}_{\bf 3}}(\tau)$.
The modular forms of weight $k=4$ have dimension $d_4=5$.
They decompose to ${\bf 3}$, ${\bf 1}$, and ${\bf 1}'$ and are written explicitly by 
\begin{align}
&{ Y^{\rm (4)}_{\bf 3}}(\tau)=
\begin{pmatrix}
Y_1^2-Y_2Y_3  \\
Y_3^2-Y_1Y_2 \\
Y_2^2-Y_1Y_3
\end{pmatrix}\,,
\nonumber\\
&Y_{\bf 1}^{(4)}=Y_1^2+2Y_2Y_3\,, \qquad\qquad 
Y_{\bf 1'}^{(4)}=Y_3^2+2Y_1Y_2\,. 
\label{Y4}
\end{align}
The modular form corresponding to the non-trivial singlet ${\bf 1}''$ 
vanishes identically $Y_{\bf 1''}^{(4)}=Y_2^2+2Y_1Y_3 =0$ \cite{Feruglio:2017spp}.
Also the modular form corresponding to ${\bf 3}_a$ vanishes.

Similarly, modular forms of higher weights are constructed (see, e.g., Refs.~\cite{Zhang:2019ngf,Lu:2019vgm}).
The modular forms of weight $k=6$ have dimension $d_6=7$.
They decompose to ${\bf 3} + {\bf 3}+{\bf 1}$,  and are written explicitly by 
\begin{align}
&{ Y^{\rm (6)}_{\bf 3,1}}(\tau)=Y_{\bf 1}^{(4)} { Y^{\rm (2)}_{\bf 3}}(\tau)=(Y_1^2+2Y_2Y_3)
\begin{pmatrix}
Y_1  \\
Y_2 \\
Y_3
\end{pmatrix}\,, \notag\\
&{ Y^{\rm (6)}_{\bf 3,2}}(\tau)=Y_{\bf 1'}^{(4)} { Y^{\rm (2)}_{\bf 3}}(\tau)=(Y_3^2+2Y_1Y_2)
\begin{pmatrix}
Y_3  \\
Y_1 \\
Y_2
\end{pmatrix}\,,
\nonumber\\
&{ Y^{\rm (6)}_{\bf 1}}(\tau)=({ Y^{\rm (4)}_{\bf 3}}{ Y^{\rm (2)}_{\bf 3}})_{\bf 1}=
Y_1^3+Y_2^3+Y_3^3-3Y_1Y_2Y_3\,, 
\label{Y6}
\end{align}
where $({ Y^{\rm (4)}_{\bf 3}}{ Y^{\rm (2)}_{\bf 3}})_{\bf 1}$ is the 
trivial singlet projection of the tensor product ${ Y^{\rm (4)}_{\bf 3}}{ Y^{\rm (2)}_{\bf 3}}$.

The modular forms of weight $k=8$ have dimension $d_8=9$.
They decompose to ${\bf 3} + {\bf 3}+{\bf 1}+{\bf 1}'+{\bf 1}''$,  and are written explicitly by 
\begin{align}
&{ Y^{\rm (8)}_{\bf 3,1}}(\tau)=Y_{\bf 1}^{(4)} Y^{\rm (4)}_{\bf 3}=(Y_1^2+2Y_2Y_3)
\begin{pmatrix}
Y_1^2-Y_2Y_3  \\
Y_3^2-Y_1Y_2 \\
Y_2^2-Y_1Y_3
\end{pmatrix}\,, \notag
\\
&{ Y^{\rm (8)}_{\bf 3,2}}(\tau)=Y_{\bf 1'}^{(4)} Y^{\rm (4)}_{\bf 3}=(Y_3^2+2Y_1Y_2)
\begin{pmatrix}
Y_2^2-Y_1Y_3  \\
Y_1^2-Y_2Y_3 \\
Y_3^2-Y_1Y_2
\end{pmatrix}\,,   \notag
\\
&{ Y^{\rm (8)}_{\bf 1}}(\tau)= ({ Y^{\rm (4)}_{\bf 1}})^2=(Y_1^2+2Y_2Y_3)^2\,, \notag \\
&{ Y^{\rm (8)}_{\bf 1'}}(\tau)= { Y^{\rm (4)}_{\bf 1}}  { Y^{\rm (4)}_{\bf 1'}} =(Y_1^2+2Y_2Y_3)(Y_3^2+2Y_1Y_2)\,,  \notag \\
&{ Y^{\rm (8)}_{\bf 1''}}(\tau)= { Y^{\rm (4)}_{\bf 1'}}  { Y^{\rm (4)}_{\bf 1'}} =(Y_3^2+2Y_1Y_2)^2\,  .
\label{Y8}
\end{align}
Note that the non-trivial singlet ${\bf 1}''$ appears when the weight $k=8$.

The modular forms of weight $k=10$ have dimension $d_{10}=11$.
They decompose to ${\bf 3} + {\bf 3}+{\bf 3}+ {\bf 1}+{\bf 1}'$,  and are written explicitly by 
\begin{align}
&{ Y^{\rm (10)}_{\bf 3,1}}(\tau)=Y_{\bf 1}^{(8)} { Y^{\rm (2)}_{\bf 3}}(\tau)=(Y_1^2+2Y_2Y_3)^2
\begin{pmatrix}
Y_1  \\
Y_2 \\
Y_3
\end{pmatrix}\,,  \notag
\\
&{ Y^{\rm (10)}_{\bf 3,2}}(\tau)=Y_{\bf 1''}^{(8)} { Y^{\rm (2)}_{\bf 3}}(\tau)=(Y_3^2+2Y_1Y_2)^2
\begin{pmatrix}
Y_2  \\
Y_3 \\
Y_1
\end{pmatrix}\,, \notag
\\
&{ Y^{\rm (10)}_{\bf 3,3}}(\tau)=Y_{\bf 1'}^{(8)} { Y^{\rm (2)}_{\bf 3}}(\tau)=(Y_1^2+2Y_2Y_3)(Y_3^2+2Y_1Y_2)
\begin{pmatrix}
Y_3  \\
Y_1 \\
Y_2
\end{pmatrix}\,,  \notag
\\
& Y^{\rm (10)}_{\bf 1}(\tau)= Y^{\rm (4)}_{\bf 1}Y^{\rm (6)}_{\bf 1}
=(Y_1^2+2Y_2Y_3)(Y_1^3+Y_2^3+Y_3^3-3Y_1Y_2Y_3)\,, \notag  \\
&Y^{\rm (10)}_{\bf 1'}(\tau)= Y^{\rm (4)}_{\bf 1'}Y^{\rm (6)}_{\bf 1}
=(Y_3^2+2Y_1Y_2)(Y_1^3+Y_2^3+Y_3^3-3Y_1Y_2Y_3)\, .
\label{Y10}
\end{align}

\begin{table}[ht]
\centering
\begin{tabular}{|c|c|c|} \hline
$k$ &$d_k$ & $A_4$ representations \\ \hline
2 & 3& ${\bf 3}$ \\ \hline
4 & 5& ${\bf 3}$ + ${\bf 1}$ + ${\bf 1}'$ \\ \hline
6 & 7& ${\bf 3}$ + ${\bf 3}$ + ${\bf 1}$ \\ \hline
8 & 9& ${\bf 3}$ + ${\bf 3}$ + ${\bf 1}$ + ${\bf 1}'$ + ${\bf 1}''$ \\ \hline
10 & 11 & ${\bf 3}$ + ${\bf 3}$ + ${\bf 3}$ + {\bf 1} + ${\bf 1}'$ \\ \hline
\end{tabular}
\caption{ $A_4$ representations for each weight $k$.}
\label{tab:A4-rep}
\end{table} 

Table \ref{tab:A4-rep} shows the $A_4$ representations for the modular forms 
of each weight $k$.

\section{4D low-energy effective field theory from higher dimensional theory}
\label{sec:high-D}

We study a scenario to derive 4D modular flavor symmetric models 
from $(4+d)$-dimensional theory by compactification. 
We assume that the modulus $\tau$ describes geometrical characters of 
$d$-dimensional compact space such as shape, 
although the compact space may have other moduli.
On top of that, we assume that the compact space has the modular symmetry on $\tau$. 
This originates from the symmetry of $\tau$, that is, the geometric symmetry of extra-dimensional spaces. For instance, 
if extra-dimensional spaces include $T^2$ or its orbifold as
subspace, such backgrounds enjoy the 
$SL(2,\mathbb{Z})$ modular symmetry when the modulus $\tau$ is identified with the complex structure modulus or the K\"ahler modulus \cite{Kobayashi:2018rad,Kobayashi:2018bff,Ohki:2020bpo,Kikuchi:2020frp,Kikuchi:2020nxn,Kikuchi:2021ogn,Almumin:2021fbk}. Furthermore, the $Sp(2h,\mathbb{Z})$ symplectic modular symmetry also appears in toroidal orbifolds with multimoduli \cite{Baur:2020yjl} and Calabi-Yau backgrounds \cite{Strominger:1990pd,Candelas:1990pi,Ishiguro:2020nuf,Ishiguro:2021ccl}. 
Our discussion can be applied for such a compact space.
In this section, we study the modular symmetric theory without specifying the extra-dimensional space. 
Most 4D modular flavor symmetric models are constructed within the 
framework of a (global) supersymmetric theory.
Hence, we assume that our compactification preserves 4D ${\cal N}=1$ supersymmetry.

We denote coordinates of 4D spacetime and $d$-dimensional compact space 
by $x$ and $y$, respectively.
Bosonic fields $\Phi(x,y)$ and spinor fields  $\Psi(x,y)$ 
in a higher dimensional theory  
are written by KK decomposition as 
\begin{align}
&\Phi(x,y) = \sum_i\phi_i(x) \varphi_i(y) + \cdots, \notag \\
&\Psi(x,y)=\sum_i\psi_i(x) \chi_i(y) + \cdots .
\end{align}
Bosonic fields $\Phi(x,y)$ correspond to scalars or vectors in $(4+d)$ dimensions, 
but vector fields with vector indices along extra dimensional space 
are 4D scalars.
The first terms on the RHSs are massless  modes (zero-modes) and the others are massive modes.
Here, we focus on massless modes.
In general, there is more than one zero-mode, which is labeled by the index $i$.
The zero-mode index $i$ corresponds to the flavor index in the 4D low energy effective field theory.
Hereafter, we often omit this index $i$.

It is  noted that 
$\phi(x)$ and $\psi(x)$ are 4D fields, while 
$\varphi(y)$ and $\chi(y)$ are wavefunctions in extra dimensions.
The wavefunctions $\varphi(y)$ and $\chi(y)$ depend on the modulus $\tau$, namely 
metric deformations of extra-dimensional space. 
Since this theory is assumed to be modular symmetric, 
these wavefunctions are modular forms \footnote{
For example, wavefunctions in magnetized D-brane models are modular forms 
\cite{Kobayashi:2018rad,Kobayashi:2018bff,Ohki:2020bpo,Kikuchi:2020frp,Kikuchi:2020nxn,
Kikuchi:2021ogn,Almumin:2021fbk}.}. 
Thus, for fixed modular weight $k$, $\Gamma_N$ representations of  
wavefunctions $\varphi(y)$ and $\chi(y)$ are constrained as in the previous section.
For example, for $N=3$, the  wavefunctions $\varphi(y)$ and $\chi(y)$ of weight $k=2$ 
are only the $A_4$ triplet, but not singlets.
When $k=4$, wavefunctions $\varphi(y)$ and $\chi(y)$ can correspond to 
either of ${\bf 3}$, ${\bf 1}$, ${\bf 1}'$, but not ${\bf 1}''$.
The singlet ${\bf 1}''$ can appear for $k=8$ and higher weights.
Hence, we have constraints on weights $k$ and $\Gamma_N$ representations 
in a higher dimensional theory. 
Note that the value of level $N$ depends on models in a higher-dimensional theory. 
For instance, the $\Gamma_3$ representations appear in twisted modes of the heterotic 
$T^2/\mathbb{Z}_3$ orbifold \cite{Ferrara:1989qb,Lerche:1989cs,Lauer:1990tm}. Other representations are also possible on heterotic 
orbifolds and magnetized D-brane models. The value of $N$ depends on the geometric 
structure of extra-dimensional spaces as well as background sources. It is important 
to reveal the geometric meaning of $N$, but we leave these issues for future research.

For fixed weight $k$, there are $d_k(\Gamma(N))$ dimensions of modular forms 
as shown in Table \ref{tab:form-dim}.
Hereafter, we assume that not all modular forms, but 
one or more irreducible representations 
of $\Gamma_N$ appear as zero-mode wavefunctions of matter fields.
Such a projection from $d_k(\Gamma(N))$ dimensions to 
irreducible representations would be possible by imposing certain 
boundary conditions.
In Appendix \ref{appendix:A}, we show an example to project out some of 
reducible representations in zero-mode wavefunctions so as to 
obtain irreducible representations.
Here, we show that such projections can be consistent with 
the modular symmetry.
We assume that a wavefunction $\varphi_1(y)$ corresponding to 
$d_k$-dimensional modular forms of $\Gamma(N)$ satisfies 
a zero-mode equation with a specific boundary condition.
When those $d_k$-dimensional wavefunctions are a reducible 
representation of $\Gamma_N$, the unitary matrix $\rho(\gamma)_{ij}$ 
is represented by 
\begin{eqnarray}
\rho(\gamma)_{ij} 
\left(
\begin{array}{c}
\varphi_1(y) \\
\vdots \\
\vdots
\end{array}
\right)=\left(
\begin{array}{ccc}
\rho^{(1)}(\gamma)_{ij} &  & \\
 & \rho^{(2)}(\gamma)_{ij} & \\
 & & \ddots 
\end{array}
\right)
\left(
\begin{array}{c}
\varphi_1(y) \\
\vdots \\
\vdots
\end{array}
\right).
\end{eqnarray}
That implies that the modular transformation of $\varphi_1(y)$ is closed 
in the irreducible representation corresponding to 
$\rho^{(1)}(\gamma)_{ij}$, but does not transform other representations such as 
$\rho^{(2)}(\gamma)_{ij}$.
Thus, it is consistent with the modular symmetry to pick up 
an irreducible representation from $d_k$-dimensional modular forms.
 
The minimal SUSY model has one pair of Higgs modes, i.e., 
the up-sector and down-sector Higgs fields.
They must be $\Gamma_N$ singlets.
Thus, it is reasonable to assign the modular weight $k=0$ to the Higgs modes.

It is natural to normalize the wavefunction of the weight $k$ as 
\begin{align}
\int d^dy\,\sqrt{g}\, |\varphi(y)|^2 = \frac{1}{(2{\rm Im}(\tau))^k},
\end{align}
with $g$ being the determinant of the metric of extra-dimensional space. 
This normalization is consistent with the modular symmetry. 
Indeed, the left-handed side of the above expressions transforms as
\begin{align}
    \int d^d y \,\sqrt{g}\, |\varphi(y)|^2 &\rightarrow |c\tau +d|^{2k}  \int d^d y\,\sqrt{g}\, |\varphi(y)|^2,
\end{align}
taking into account the modular transformation of $\varphi(y)$:
\begin{align}
    \varphi(y) \rightarrow \rho(\gamma) (c\tau +d)^k \varphi(y).
    \label{eq:modular_matter}
\end{align}
Note that $\int d^dy \sqrt{g}$ is invariant under the modular transformation, i.e., 
the coordinate transformation. Thus, it is consistent with the modular transformation 
of $\tau$:
\begin{align}
    (2{\rm Im}(\tau))^{-k} &\rightarrow |c\tau +d|^{2k} (2{\rm Im}(\tau))^{-k}.    
\end{align}
We start with the following canonical kinetic term,
\begin{align}
\partial_M \Phi^* \partial^M \Phi ,
\end{align}
in a higher dimensional theory.\footnote{Note that other higher-dimensional higher-derivative and interaction terms will provide the correction terms in the K\"ahler potential, but they would be suppressed by the compactification scale.}
Then, we carry out the dimensional reduction by the use of the above normalization 
so as to obtain the following K\"ahler potential of matter fields:
\begin{align}
K =  \frac{1}{(2{\rm Im}(\tau ))^k}|\phi(x)|^2.
\label{eq:Kahler}
\end{align}
We require that the 4D effective field theory is invariant under the modular transformation.
The 4D matter fields must have the modular weights $-k$, which have
opposite signs compared with the wave function weights $k$.
Here and hereafter, we use the notation that 
the same letter is used for both the superfield and its lowest scalar component.

Next, we study the Yukawa coupling terms in the superpotential.
For example, suppose that the Yukawa coupling terms in the 4D effective theory 
originate from the following terms in a higher dimensional theory:
\begin{align}
y \bar \Psi_e \Phi^*_{H} \Psi_L ,
\end{align}
where $\Phi^*_{H}$ is the higher dimensional field corresponding to 
the 4D down-sector Higgs field $H_d$, and 
$\Psi_{L},\Psi_e$ are higher dimensional fields corresponding to 
left-handed and right-handed leptons in the 4D effective theory.
Then, we integrate the extra dimensions $y$ so as to 
derive the Yukawa coupling terms in the 4D superpotential,
\begin{align}
W =Y_e(\tau)LH_d e^c.
\end{align}
Here, the 4D Yukawa coupling $Y_e(\tau)$ is obtained by 
\begin{align}
Y_e(\tau) = y\int d^dy~\sqrt{g}~ \chi_{e^c}(y) ~\chi_L(y) ~\varphi^*_H(y).
\end{align}
The 4D modes $L$ and $e^c$ have modular weights $-k_L$ and $-k_e$, respectively, 
while $H_d$ has vanishing modular weight.
The 4D Yukawa coupling $Y_e(\tau)$ has modular weight $k_L + k_e$ 
because the product of wavefunctions in the extra dimension has weight $k_L + k_e$.
Then, the above superpotential is invariant under the modular transformation. 
Indeed, the modular transformation of matter fields (\ref{eq:modular_matter}) 
induces the correct modular transformation of the Yukawa couplings:
\begin{align}
    Y_e(\tau) \rightarrow (c\tau+d)^{k_L+k_e}\rho(\gamma) Y_e(\tau).
\end{align}

The same result can be derived by another way as follows.
The product of wavefunctions $\chi_{e^c}(y) ~\chi_L (y) $ can be 
expanded by all the KK wavefunctions $\Phi_H$,
\begin{align}
\chi_L(y)\chi_{e^c}(y) = Y_e(\tau) \varphi_H(y) + \cdots,
\end{align}
since all the KK wavefunctions are a complete set.
The first term on the RHS corresponds to the massless mode, while 
the others are massive modes 
\footnote{In specific higher dimensional theories, the production of massless 
wavefunctions can be expanded only by massless modes \cite{Cremades:2004wa,Abe:2009dr,Honda:2018sjy}.}.
The expansion coefficient $Y_e(\tau)$ corresponds to the 4D Yukawa coupling.
Both sides must have the same modular weight, and $\varphi_H$ has vanishing weight.
Thus, the 4D Yukawa coupling $Y_e(\tau)$ has modular weight $k_L+ k_e$.

Thus, we can derive the 4D modular flavor symmetric model 
from a higher dimensional theory.
In this scenario, we have the constraint on combinations between 
modular weights and $\Gamma_N$ representations of matter fields, 
although one has assigned modular weights and $\Gamma_N$ representations 
to matter fields without such a constraint in modular flavor models, which have 
been constructed so far.
For example, one can not assign odd weights to matter fields 
in modular $\Gamma_N$ flavor models. 
The modular $\Gamma_N'$ flavor symmetry is then required 
to assign odd weights to matter fields.
In the next section, we show  $A_4$ modular flavor models 
as illustrating examples.

\section{Examples in $A_4$ modular flavor models}
\label{sec:model}

In the previous section, we have studied a scenario to derive 
4D modular flavor symmetric models from higher dimensional theory.
In this scenario, we have the constraint on combinations of 
modular weights and $\Gamma_N$ representations for matter fields.
For example, matter fields must have even modular weights 
in $A_4$ modular flavor models.
The matter fields with modular weight $-k=-2$ must be 
$A_4$ triplet, but other representations can not be allowed.
The non-trivial $A_4$ singlet ${\bf 1}''$ can not be assigned to the matter fields 
with modular weight $-k=-2, -4, -6$, but can be the matter fields with weight $-k=-8$. 
We present $A_4$ models.

In many $A_4$ models, three generations of lepton doublets $L$ are 
assigned to the $A_4$ triplet ${\bf 3}$, 
and three generations of  right-handed charged leptons $e^c$ are assigned to 
three $A_4$ singlets, ${\bf 1}, {\bf 1}'', {\bf 1}'$.
In order to use such assignments of $A_4$ representations, 
we study the model that three generations of lepton doublets $L_i$ have  modular weight $-k=-2$ and three generations of right-handed charged leptons 
$e^c_i$ have modular weight $-k=-8$.
Such an assignment is summarized in Table \ref{tab:A4-model}.

\begin{table}[h]
\centering
\begin{tabular}{|c|c|c|c|} \hline
   &$L_i$ & $e_i^c$& $H_d$  \\ \hline \hline
$SU(2)$ & ${\bf 2}$ & ${\bf 1}$ & ${\bf 2}$ \\ \hline
$A_4$ & ${\bf 3}$ & ${\bf 1},\, {\bf 1}'',\, {\bf 1}'$ & ${\bf 1}$ \\ \hline
$k$ & $-2$ & $-8$ & $0$  \\ \hline
\end{tabular}
\caption{ Assignment of $A_4$ representations and weights.}
\label{tab:A4-model}
\end{table}

The $A_4$ modular invariant superpotential relevant to the lepton sector can be written by 
\begin{align}
W= \sum_{\bf r}\alpha_{\bf r} (Y^{(10)}_{\bf r}L)_{1}H_d\,e^c_1 +\sum_{\bf r}\beta_{\bf r}(Y^{(10)}_{\bf r}L)_{1'}H_d\,e^c_{1''} \notag \\
+ \sum_{\bf r}\gamma_{\bf r} (Y^{(10)}_{\bf r}L)_{1''}H_d\,e^c_{1'} 
+ \sum_{\bf r} g_{\bf r} \frac{Y^{(4)}_{\bf r}}{\Lambda}LHLH.
\end{align}
We set $\alpha_{\bf r}=\beta_{\bf r}=\gamma_{\bf r}=0$ except ${\bf r}=({\bf 3,1})$ of Eq.\eqref{Y10}, and the dimensionful parameter $\Lambda$ is set to obtain the correct scale of neutrino masses. On the other hand,
non-vanishing $g_{\bf r}$'s are given for ${\bf r}={\bf 3},\,{\bf 1},\,{\bf 1'}$
of Eq.\eqref{Y4}.
Then, this superpotential is quite similar to the one in Ref.~\cite{Okada:2020brs}.
Indeed, we set 
\begin{align}
&\tau = 0.0796+1.0065\, i\,, \qquad\quad g_{\bf 1}/g_{\bf 3}=0.124\,, 
\quad\qquad g_{\bf 1'}/g_{\bf 3}=-0.802\,, 
\notag \\
&\alpha_{\bf 3,1}/\gamma_{\bf 3,1}=6.82 \times 10^{-2}, \qquad 
\beta_{\bf 3,1}/\gamma_{\bf 3,1}=1.02 \times 10^{-3} ,
\end{align}
so as to realize realistic values of charged lepton mass ratios
and   neutrino mass squared differences.
The obtained mixing angles are 
\begin{align}
\sin^2 \theta_{12} = 0.294\,, \qquad \sin^2 \theta_{23}= 0.563\,, 
\qquad \sin^2 \theta_{13}=0.0226\,,
\end{align}
which are within a $1\,\sigma$ error bar of observed values \cite{Esteban:2020cvm}.
Thus, we can construct the modular flavor symmetric models, which are consistent 
with our scenario and can derive realistic results.

Although the above model is a simple model, 
we may be able to study other assignments consistent with our scenario.
For example, we assign $A_4$ representations and modular weights to 
three generations of $e^c_i$ as 
${\bf 1}$ (weight $-4$), ${\bf 1}''$ (weight $-8$), and ${\bf 1}'$ (weight $-4$), while 
we use the same assignment for $L_i$ and $H_d$.
Then, the modular $A_4$ invariant superpotential can be written by 
\begin{align}
W= \sum_{\bf r}\alpha_{\bf r} (Y^{(6)}_{\bf r}L)_{1}H_d e^c_1 +\sum_{\bf r}\beta_{\bf r}(Y^{(10)}_{\bf r}L)_{1'}H_de^c_{1''} \notag \\
+ \sum_{\bf r}\gamma_{\bf r} (Y^{(6)}_{\bf r}L)_{1''}H_de^c_{1'} 
+ \sum_{\bf r} g_{\bf r} \frac{Y^{(4)}_{\bf r}}{\Lambda}LHLH.
\end{align}
We also set $\alpha_{\bf r}=\beta_{\bf r}=\gamma_{\bf r}=0$ except 
${\bf r}=({\bf 3,1})$ of Eqs.\eqref{Y6} and \eqref{Y10}.
By using proper values of the parameters, 
we can realize almost the same results of the lepton masses and mixing angles as the previous model.
In this model, three generations of $e^c_i$ have two different modular weights, $-4$ and $-8$.
Thus, these three generations may originate from not a single field $\Phi(x,y)$ in a higher dimensional 
theory,  but at least two fields $\Phi(x,y)$ and $\Phi'(x,y)$ where 
one field corresponds to modular weight $-4$ and the other corresponds to weight $-8$.

As another model, we assign three generations of $e^c_i$ as 
${\bf 1}$ (weight $-4$), ${\bf 1}'$ (weight $-4$), and ${\bf 1}'$ (weight $-4$), while 
we use the same assignment for $L_i$ and $H_d$.
Then, the modular $A_4$ invariant superpotential can be written by 
\begin{align}
W= \sum_{\bf r}\alpha_{\bf r} (Y^{(6)}_{\bf r}L)_{1}H_de~c_1 +\sum_{\bf r}\beta_{\bf r}(Y^{(6)}_{\bf r}L)_{1''}H_de^c_{1'} \notag \\
+ \sum_{\bf r}\gamma_{\bf r} (Y^{(6)}_{\bf r}L)_{1''}H_de^c_{1'} 
+ \sum_{\bf r} g_{\bf r} \frac{Y^{(4)}_{\bf r}}{\Lambda}LHLH.
\end{align}
Taking $\alpha_{\bf r}=\gamma_{\bf r}=0$ except 
${\bf r}=({\bf 3,1})$  and 
$\beta_{\bf r}=0$ except  ${\bf r}=({\bf 3,2})$ of Eq.\eqref{Y6},
we can also realize almost the same results of the lepton masses and mixing angles as the previous model. 
Since three generations of $e^c_i$ have the same modular weight in this model, 
they can originate from a single field $\Phi(x,y)$ in a higher dimensional theory.
In this model, two modes have the same $A_4$ representation ${\bf 1}'$ and the same weight $-4$.
They may have different properties on boundary conditions in extra dimensions, e.g., 
$Z_N$ twist eigenvalues.
Alternatively, we assign three generations of $e^c_i$ as 
${\bf 1}$ (weight $-4$), ${\bf 1}'$ (weight $-4$), and ${\bf 1}$ (weight $-4$).

These models are consistent with our scenario and can lead to realistic lepton masses and 
mixing angles.
One of the important issues is to study their difference in particle phenomenology, i.e., 
how to distinguish these models.
The first model in Table \ref{tab:A4-model} and the second model have 
different modular weights for the matter fields with the representation 
$\{{\bf 1}, {\bf 1}'\}$.
Here, we give a comment on the phenomenological difference due to the modular weights.

Within the framework of supergravity theory, soft scalar masses $m_i$ with the 
moduli-dependent K\"ahler metric 
$K_{i \bar i}$ are given as \cite{Kaplunovsky:1993rd}
\begin{eqnarray}
m_i= m_{3/2}^2-\sum_X |F^X|^2 \partial_X \partial_{\bar X} \ln K_{i \bar i}\,,
\end{eqnarray}
when $F$-terms $F^X$ of the moduli $X$ develop their vacuum expectation values.
Suppose that the $F$-term $F^\tau$ of the modulus $\tau$ develops its vacuum expectation value.
Then, the K\"ahler metric in Eq.(\ref{eq:Kahler}) leads to the soft masses \cite{Kobayashi:2021jqu}
\begin{eqnarray}
m_i^2= m_{3/2}^2-k_i \frac{|F^\tau|^2}{(2{\rm Im}\tau)^2} \,.
\end{eqnarray}

The first model  in Table \ref{tab:A4-model} leads to 
degenerate soft masses in three generations of right-handed leptons as well as 
left-handed leptons.
In the second model, the right-handed lepton with the representation ${\bf 1}''$ 
has a modular weight different from the other.
Thus, their slepton masses are not degenerate.

Similarly, 
in the third model, three generations of slepton masses are degenerate.
We may have a difference between the first and third models 
in higher-dimensional operators in the SMEFT \cite{Kobayashi:2021pav}.

\section{Conclusion}
\label{sec:conclusion}

We have studied the scenario to derive 4D modular flavor symmetric models 
from a higher dimensional theory. 
In our scenario, wavefunctions in extra dimensions are modular forms. 
That leads to the constraints on combinations between modular weights and $\Gamma_N$ ($\Gamma_N'$) representations, which have not been considered in the bottom-up approach. 
As illustrating examples, we have shown explicit $A_4$ models, taking into account the constraints from a higher dimensional theory. 
It is found that realistic results on lepton masses and mixing angles are realized. 
Our discussions can also be applied to the quark sector, though we do not discuss them here. 
We can extend them to other $\Gamma_N$ models and their covering groups.
Further studies along our scenario would be important to 
connect 4D flavor models with higher dimensional theory such as the superstring theory.



\vspace{1.5 cm}
\noindent
{\large\bf Acknowledgement}\\

This work was supported by JSPS KAKENHI Grants No. JP19J00664 (H. O.), 
No. JP20K14477 (H. O.), No. JP20J20388 (H. U.), and No. JP21K13923 (K. Y.), and JST SPRING Grant No. JPMJSP2119 (S. K.).

\appendix
\section*{Appendix}


\section{Projection by boundary condition}
\label{appendix:A}

Here, we show an example to project wavefunctions with 
$\Gamma_N$ reducible representations to 
an irreducible one by imposing further boundary conditions.

Suppose that the following five wavefunctions satisfy the same zero-mode equations 
in some compactification with the complex coordinate $z=y_1+\tau y_2$, 
which may have other dimensional coordinates,
\begin{align}
\begin{array}{l}
\chi_1(z,\tau) \equiv (\psi^{0,2}(z,\tau))^4 + (\psi^{1,2}(z,\tau))^4, \\
\chi_2(z,\tau) \equiv 2\sqrt{3}  (\psi^{0,2}(z,\tau))^2 (\psi^{1,2}(z,\tau))^2,\\
\chi_3(z,\tau) \equiv (\psi^{0,2}(z,\tau))^4 - (\psi^{1,2}(z,\tau))^4, \\
\chi_4(z,\tau) \equiv 2 \left( (\psi^{1,2}(z,\tau))^3 \psi^{1,2}(z,\tau) + \psi^{0,2}(z,\tau) (\psi^{1,2}(z,\tau))^3 \right), \\
\chi_5(z,\tau) \equiv 2 \left( (\psi^{1,2}(z,\tau))^3 \psi^{1,2}(z,\tau) - \psi^{0,2}(z,\tau) (\psi^{1,2}(z,\tau))^3 \right), \\
\end{array}
\label{eq:X15}
\end{align}
where 
\begin{align}
\psi^{j,M}(z,\tau) \equiv \left(\frac{M}{{\cal A}^2}\right)^{1/4} e^{\pi iMz\frac{{\rm Im}z}{{\rm Im}\tau}} \vartheta
\begin{bmatrix}
\frac{j}{M} \\ 0
\end{bmatrix}
(Mz, M\tau), \quad j\in \mathbb{Z}/M\mathbb{Z},
\end{align}
and $\vartheta$ denotes the Jacobi-theta function defined by
\begin{align}
\vartheta
\begin{bmatrix}
a \\ b
\end{bmatrix}
(\nu,\tau)
= \sum_{\ell \in \mathbb{Z}} e^{\pi i (a+\ell) \tau} e^{2\pi i (a+\ell)(\nu + b)}.
\label{theta}
\end{align}

These wavefunctions transform each other under the modular symmetry.
Under the $S$ transformation with $z \to -z/\tau$, 
these wavefunctions transform 
\begin{align}
\chi_i(z,\tau) \to (-\tau)^2\rho(S)_{ij}\chi_j(z,\tau),
\end{align}
where
\begin{eqnarray}
\rho(S)_{ij} 
=\left(
\begin{array}{cc}
\rho^{(1)}(S)_{ij} & 0 \\
0 & \rho^{(2)}(S)_{ij} 
\end{array}
\right),
\end{eqnarray}
\begin{align}
\rho^{(1)}(S) = - \frac{1}{2}
\begin{pmatrix}
1 & \sqrt{3} \\
\sqrt{3} & -1
\end{pmatrix}, \quad
\rho^{(2)}(S) = -
\begin{pmatrix}
0 & 1 & 0 \\
1 & 0 & 0 \\
0 & 0 & 1
\end{pmatrix} .
\end{align}
The above behavior implies that the above wavefunctions have modular weight $2$.
Under the $T$ transformation, these wavefunctions transform 
\begin{align}
\chi_i(z,\tau) \to \rho(T)_{ij}\chi_j(z,\tau),
\end{align}
where
\begin{eqnarray}
\rho(T)_{ij} 
=\left(
\begin{array}{cc}
\rho^{(1)}(T)_{ij} & 0 \\
0 & \rho^{(2)}(T)_{ij}  
\end{array}
\right),
\end{eqnarray}
\begin{align}
\rho^{(1)}(T) =
\begin{pmatrix}
1 & 0 \\
0 & -1
\end{pmatrix},
\quad
\rho^{(2)}(T) = 
\begin{pmatrix}
1 & 0 & 0 \\
0 & 0 & i \\
0 & i & 0
\end{pmatrix}.
\end{align}
$\rho(S)$ and $\rho(T)$ are representations of $\Gamma_4 \simeq S_4$.
In particular, $\chi_i$ are reducible representations.
$(\chi_1,\chi_2)$ correspond to the doublet ${\bf 2}$ of $S_4$, while 
$(\chi_3,\chi_4,\chi_5)$ correspond to the triplet ${\bf 3}'$.

In addition to the above compactification, we impose 
further boundary conditions.
We study the shifts of the coordinate, 
\begin{align}
z \rightarrow z+(m+n\tau)/2, \qquad (m,n)=(1,0), (0,1), (1,1).
\end{align}
The modes $(\chi_1,\chi_2)$ are invariant under 
all of these shifts.
On the other hand, the modes $(\chi_3, \chi_4,\chi_5)$ transform
\begin{align}
\chi_i \to e^{\pi i Q^i_{(m,n)}}\chi_i,
\end{align}
where 
\begin{align}
Q^3_{(m,n)}=(0,1,1), \quad Q^4_{(m,n)}=(1,0,1), \quad Q^5_{(m,n)}=(1,1,0), 
\end{align}
for $(m,n)=(1,0), (0,1), (1,1)$, respectively.
Thus, if we require the shift invariance of wavefunctions, 
we can project the five wavefunctions, $\chi_i$ with the representations 
${\bf 2}+{\bf 3}'$ to the irreducible representation ${\bf 2}$, $(\chi_1,\chi_2)$.
(For shift invariance, see Refs.~\cite{Kikuchi:2020frp,Kikuchi:2021ogn,Fujimoto:2013xha}.)

\clearpage

\end{document}